\documentclass[twocolumn,letter]{jpsj3}
\usepackage{bm}

\title{Strong Anisotropy in Spin Suceptibility of Superfluid ${}^3$He-B 
Film Caused by Surface Bound States}

\author{Yasushi \textsc{Nagato}, 
Seiji \textsc{Higashitani} and Katsuhiko \textsc{Nagai} }

\inst{Graduate School of Integrated Arts and Sciences, Hiroshima
University,
Kagamiyama 1-7-1, Higashihiroshima 739-8521, Japan
}

\abst{Spin susceptibility of  superfluid ${}^3$He-B film with specular
surfaces
is calculated. It is shown that, when the magnetic field is applied
in a direction perpendiculr to the film, the suseptibility is
significantly
 enhanced by the contribution from the surface bound states.
No such enhancement is found for the magnetic field parallel to the
film. A simplified model with spatially constant order parameter is used
to elucidate the magnetic properties of the surface bound states.
The Majorana nature of the zero energy bound state is also mentioned.
}

\kword{Superfluid ${}^3$He-B Film, Anisotropic Spin Susceptibility, Surface Bound
State}

\usepackage{bm}
\begin{document}
\maketitle
 Existence of low-energy midgap states in the vicinity of the surface and/or interface
is a universal feature of unconventional superconductors and
superfluids.\cite{bz,hn,Hu,tk}
The surface bound states govern the transport properties of 
superconductors and superfluids, 
since the system always communicates with its environment through the surface. 
For example, 
the zero-bias conductance peak (ZBCP) in the tunneling spectrum of 
high $T_c$ superconductors\cite{tk,capg,kt} has been ascribed to the zero-energy bound states.
ZBCP has been also observed in other superconductors, such as Sr$_2$RuO$_4$\cite{Laube}, 
$\kappa$-(BEDT-TTF)$_2$Cu[N(CN)$_2$]Br\cite{Ichimura}, UBe$_{13}$\cite{Ott}, and CeCoIn$_5$\cite{Wei2}, and 
was regarded as evidence of their unconventionality. 

The surface bound states 
in $p$-wave pairing superfluid ${}^3$He\cite{bz,hn,Zhang,roughp,vs} 
have not been observed until recently, 
because of the lack of an appropriate probe for the neutral
superfluid. It has been recently reported, \cite{nagato2007,Nagai2008,Aoki,Saitoh,Murakawa} 
however,
that the
transverse acoustic impedance $Z$ provides useful information on the
density of states of the surface bound states in superfluid ${}^3$He-B.
The surface bound states in this system have recently attracted much attention
as Majorana fermion surface states which characterize the topological
symmetry of the bulk BW state\cite{Qi,Schnyder,Roy,Kitaev,Chung,Volovik}.
Recently Chung and Zhang\cite{Chung} suggested that, 
when the spin quantization axis
is taken parallel to the surface, the low energy behavior of the field
operators looks like that of Majorana Fermions, as a result, the 
magnetism of surface bound states is Ising like  polarized only
in the direction of surface normal.  It follows that
the susceptibility will become anisotropic. They proposed to
detect
this anisotropy by measuring the spin relaxation rate of an electron
that forms a bubble trapped near the ${}^3$He liquid free surface.

In this letter, we discuss the spin susceptibility of superfluid ${}^3$He-B
film. The properties of films with width of several coherence lengths
are expected to be dominated by the surface properties. We show that
the susceptibility indeed shows anisotropy. When the magnetic field 
is perpendicular to the film, the susceptibility is significantly enhanced.
The susceptibility of thin films  even exceeds the Pauli
susceptibility.
However, the spin susceptibility at $T=0$ remains finite and does not
diverge as Ising spin systems. 
When the magnetic field is parallel to the film, no enhancement of the
susceptibility is found.  

Let us briefly review the surface bound states.
We consider a superfluid ${}^3$He-B
filling $z>0$ domain with a specularly reflecting plane
surface located at $z=0$.
The Hamiltonian of the system in 4-dimensional Nambu representation is
\begin{equation}
 \mathcal{H}=\frac{1}{2}\int d\bm{r}d\bm{r}'\hat{\Psi}^\dagger({\bm{r}})
\mathcal{E}(\bm{r},\bm{r}')\hat{\Psi}(\bm{r}'),\label{hamiltonian}
\end{equation}
where
\begin{equation}
 \hat{\Psi}(\bm{r})=\begin{pmatrix} \psi_\uparrow \\ 
 \psi_\downarrow\\ \psi^\dagger_\uparrow \\ \psi^\dagger_\downarrow
\end{pmatrix}
\end{equation}
is the Fermion field operator in the Nambu representation. The energy matrix
$\mathcal{E}(\bm{r},\bm{r}')$ in the presence of magnetic field 
is a $4\times 4$ matrix given by
\begin{equation}
\begin{split}
\mathcal{E}(\bm{r},\bm{r}')=&
 \begin{pmatrix}
  \xi(\nabla)\delta(\bm{r}-\bm{r}') & \Delta(\bm{r},\bm{r}')\\
  \Delta^\dagger(\bm{r}',\bm{r}) & -\xi(\nabla)\delta(\bm{r}-\bm{r}')
 \end{pmatrix}\\
 &-\frac{\gamma \bm{H}}{2}\cdot 
\begin{pmatrix}
 \bm{\sigma} & \\ & -\tilde{\bm{\sigma}}\label{energy}
\end{pmatrix} \delta (\bm{r}-\bm{r}')
\end{split}
\end{equation}
with 
$\gamma$ the gyromagnetic ratio of the ${}^3$He atom  and 
$\tilde{\bm{\sigma}}$ the transpose of Pauli matrix $\bm{\sigma}$.
\def\Kp{\bm{K}_\parallel}
\def\uk{u_{k}(z)}
The order parameter of the BW state with a surface is given by
\begin{align}
\Delta(\bm{r},\bm{r}')&=\sum_{\bm{p}} e^{i\bm{p}\cdot(\bm{r}-\bm{r}')}
\Delta(z,\hat{\bm{p}}) \\
\Delta(z,\hat{\bm{p}})&=
\begin{pmatrix}
\Delta_0(z)(-\hat{p}_x+i\hat{p}_y) & \Delta_1(z)\hat{p}_z \\
\Delta_1(z)\hat{p}_z &\Delta_0(z)(\hat{p}_x+i\hat{p}_y)
\end{pmatrix}
\end{align}
where $\hat{\bm{p}}$ is a unit vector along $\bm{p}$.
We seek surface bound states solving the Bogoliubov equation
\begin{equation}
 \int d\bm{r}' \mathcal{E}(\bm{r},\bm{r}')\Psi(\bm{r}')=E\Psi(\bm{r})
\end{equation}
under the boundary condition $\Psi=0$ at $z=0$.

We assume here that $\Delta_0$ and $\Delta_1$ are constant,
for simplicity. 
Using the quasi-classical approximation
we can obtain the
bound state wave functions (Nambu amplitudes).\cite{Nagai2008}

Let us first consider the case without magnetic field.
We can find both positive and negative energy bound state
for each Fermi momentum $(\Kp, k)$, where $\Kp$ is the component
parallel to the surface and $k$ is the perpendicular component.
The bound state energies are given by $\pm \Delta_0\sin\theta$
with $\theta$  the polar angle of the Fermi momentum with respect
to the $z$ axis.\cite{roughp}
For positive energy eigen value $E=\Delta_0\sin\theta$, the
eigen function is given by  
\begin{equation}
 \Psi_{\Kp}^{(+)}(\bm{r})=e^{i\Kp\cdot\bm{r}}u(k,z)\left(\Phi_+ -
e^{i{\phi}}\Phi_-\right)\label{psiplus}
\end{equation}
and for negative energy eigen value $E=-\Delta_0\sin\theta$
\begin{equation}
 \Psi_{\Kp}^{(-)}(\bm{r})=e^{i\Kp\cdot\bm{r}}u(k,z)
\left(e^{-i{\phi}}\Phi_+  +\Phi_-\right),\label{psiminus}
\end{equation}
where $\phi$ is the azimuthal angle
of the Fermi momentum around the $z$ axis.
\noindent
Here $\Phi_\pm$ are the Nambu amplitudes given by
\begin{equation}
 \Phi_+=\begin{pmatrix} 1 \\ 0 \\ 0 \\ -i \end{pmatrix},
\qquad \Phi_-=\begin{pmatrix} 0 \\i \\ 1 \\ 0 \end{pmatrix}.
\end{equation}
It is worth noting here that $\Phi_\pm$ is the eigen vector of
spin operator $S_z$ in the Nambu representation
\begin{equation}
 S_z=\frac{1}{2}\begin{pmatrix}\sigma_z & \\ & -\sigma_z
		\end{pmatrix},\quad
S_z\Phi_\pm=\pm\frac{1}{2}\Phi_\pm.
\end{equation}
The $z$ dependence of the eigen functions is included in
\begin{equation}
 u(k,z)=u e^{-\kappa z}\sin{kz}
\end{equation}
with $\kappa=\Delta_1/v_F$ and the normalization constant $u$ is
determined so that $\Psi_{\Kp}^{(\pm)}$ are normalized.

Since all the eigen functions of $\mathcal{E}(\bm{r},\bm{r}')$ form a
complete set, we can expand the Fermion field operator $\hat{\Psi}$:
\begin{equation}
 \hat{\Psi}(\bm{r})=\sum_{\Kp} \left( \gamma_{\Kp}\Psi_{\Kp}^{(+)}(\bm{r})+
\gamma^\dagger_{-\Kp}\Psi_{\Kp}^{(-)}(\bm{r})\right)+\cdots,
\end{equation} 
where we have omitted the gapped modes. Since all the eigen functions
are
orthogonal to each other, 
we obtain
\begin{align}
 \gamma_{\Kp}=\int & d\bm{r}  e^{-i\Kp\cdot\bm{r}}u(k,z) \nonumber\\
 &\times\left(
<\Phi_+|\hat{\Psi}>-e^{-i{\phi}}<\Phi_-|\hat{\Psi}>
\right), \\
 \gamma_{\Kp}^\dagger=\int & d\bm{r}  e^{i\Kp\cdot\bm{r}}u(k,z) \nonumber\\
 &\times\left(-e^{i{\phi}}<\Phi_+|\hat{\Psi}>
+<\Phi_-|\hat{\Psi}>
\right),
\end{align}
where
\begin{align}
 <\Phi_+|\hat{\Psi}>=&\psi_\uparrow(\bm{r})+i\psi^\dagger_\downarrow(\bm{r}),
\\
<\Phi_-|\hat{\Psi}>=&(-i)\psi_\downarrow(\bm{r})+\psi^\dagger_\uparrow(\bm{r}).
\end{align}
We can show that $\gamma_{\Kp}$ and
$\gamma_{\Kp}^\dagger$ satisfy the Fermion commutation relation
\begin{equation}
 \{\gamma_{\Kp},\gamma_{\Kp'}\}=0,\qquad
  \{\gamma_{\Kp},\gamma_{\Kp'}^\dagger\}
=\delta_{\Kp.\Kp'}. 
\end{equation}

Some caution is necessary about the doubly degenerate zero energy states
which happen when $\Kp=0$. In this case,
the system is essentially the polar state\cite{hn} 
and the azimuthal angle $\phi$
is an irrelevant quantity. We can choose any linear combination of 
$\Psi_0^{(\pm)}$ as an eigen function of the zero energy state.
An example is
\begin{equation}
 \hat{\Psi}=\left(\gamma_0\Phi_-+\gamma_0^\dagger
	     \Phi_+\right)\sqrt{2}u(k_F,z)
+\mbox{nonzero energy states}.
\end{equation}
In this case, $\gamma_0, \gamma_0^\dagger$ are still Fermion operators
because $\Phi_{\pm}$ are mutually orthogonal. 
Another choice is
\begin{equation}
 \hat{\Psi}=\left(\Gamma_+\left(\Phi_-+\Phi_+\right)
+\Gamma_-\frac{1}{i}\left(\Phi_--\Phi_+\right)\right)u(k_F,z)
+\cdots
\end{equation}
with $\Gamma_+=(\gamma_0+\gamma^\dagger_0)/\sqrt{2}, 
\Gamma_-=i(\gamma_0-\gamma^\dagger_0)/\sqrt{2}$.
The new operators $\Gamma_\pm$ have a Majorana property
\begin{equation}
 \Gamma_+^\dagger=\Gamma_+,\ \  \Gamma_-^\dagger=\Gamma_-
\end{equation}
and obey the commutation relation
\begin{equation}
 \{\Gamma_+,\Gamma_+\}=\{\Gamma_-,\Gamma_-\}=1,\quad \{\Gamma_+,\Gamma_-\}=0.
\end{equation}

Now let us consider the effect by magnetic field.
To obtain the low energy spectrum, we consider  matrix
elements of $\mathcal{E}$ of Eq.~(\ref{energy})
between the eigen functions given by Eqs.~(\ref{psiplus}) 
and (\ref{psiminus}). 
It is quite interesting that only $S_z$ has a finite matrix element
between $\Psi_{\Kp}^{(+)}$ and $\Psi_{\Kp}^{(-)}$. Other spin components
$S_x, S_y$ have no matrix element at all. It implies that the 
surface bound states respond only to the magnetic field
in the direction of the surface normal. This agrees
with the recent suggestion by Chung and Zhang.\cite{Chung}
When the magnetic field is applied in the $z$-direction,
the surface bound state wave function is given by
\begin{equation}
 a \Psi_{\Kp}^{(+)}+ b \Psi_{\Kp}^{(-)}\label{magnetic_w_f}
\end{equation}
and the energy is obtained from an eigen value equation
\begin{equation}
 \begin{pmatrix}\Delta_0\sin\theta  & -\dfrac{\gamma H}{2}e^{-i\phi}\\
-\dfrac{\gamma H}{2}e^{i\phi} & -\Delta_0\sin\theta\end{pmatrix}
\begin{pmatrix} a \\ b \end{pmatrix}
=E \begin{pmatrix} a \\ b \end{pmatrix}
\end{equation}
to be 
\begin{equation}
 E=\pm \sqrt{(\Delta_0\sin\theta)^2+\left(\frac{\gamma H}{2}\right)^2}.
\end{equation}
In the $\Kp\rightarrow 0$ limit, the wave function (\ref{magnetic_w_f})
is reduced to $\Phi_-$ for the positive energy $+\frac{\gamma H}{2}$ and
to $\Phi_+$ for the negative energy $-\frac{\gamma H}{2}$.
Let us consider the spin susceptibility of the ground state. 
Contribution from the occupied negative energy bound states to the
ground state energy is given by
\begin{align}
E_0(H)
&= \frac{1}{2}\sum_{\Kp} -\sqrt{(\Delta_0\sin\theta)^2+\left(\frac{\gamma
						    H}{2}\right)^2}
\label{ground_energy}\\
&= -\frac{k_F^2}{4\pi}
\int_0^{1} d(\cos\theta)
\cos\theta\sqrt{(\Delta_0\sin\theta)^2+\left(\frac{\gamma
						  H}{2}\right)^2}\nonumber\\
&= -\frac{k_F^2}{12\pi}
\frac{1}{\Delta_0^2}\left(\left(\Delta_0^2+\left(\frac{\gamma
					    H}{2}\right)^2\right)^{3/2}-\left(\frac{\gamma |H|}{2}\right)^3\right) \nonumber\\
&\sim -\frac{k_F^2}{12\pi}\Delta_0\left(1+\frac{3}{2}
\left(\frac{\gamma H}{2\Delta_0}\right)^2+\cdots\right),
\end{align}
where the factor $1/2$ in Eq.~(\ref{ground_energy}) comes from the
prefactor
in Eq.~(\ref{hamiltonian}).
We obtain the susceptibility
\begin{equation}
 \chi_{zz}= -\frac{\partial^2}{\partial H^2}E_0(H)= \frac{\gamma^2 k_F^2}{16\pi\Delta_0}
\end{equation}
which is as large as the normal state susceptibility $\chi_N$ 
multiplied by the
width $1/\kappa=v_F/\Delta_1$ of the bound states. The susceptibility
of the surface BW state at $T=0$ is large but finite, while in the polar state
the susceptibility will diverge because $\Delta_0=0$. The magnetism of
the polar state is just Ising spin like because the bound state energy
under magnetic field splits into  $\pm\frac{\gamma H}{2}$.

\begin{figure}[h]
\includegraphics[width=8cm]{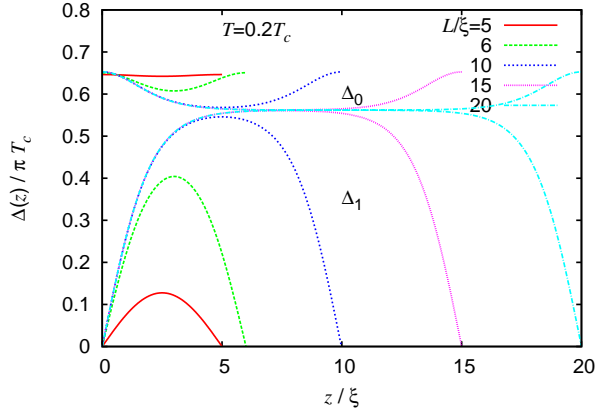}
\caption[ ]{Self consistent order parameter at $T=0.2T_c$ 
of ${}^3$He-B film with
width $L=5,6,10,15,20 \xi$. The coherence length $\xi$ is defined
by $\xi=v_F/ \pi T_c$, which is equal to 0.155$\mu$m at SVP.
}
\label{fig1}
\end{figure}

Now we turn to the susceptibility of films.
To calculate the susceptibility, we use the quasi-classical
Green function method developed for dealing with the boundary
problems.\cite{aahn,rough,roughp,higashitani_nagai} 
Within the quasi-classical Green
function theory, however, it is not straightforward to calculate the linear
response in finite systems because the quasi-classical Eilenberger
equation needs an initial condition at some point. We instead start
from the linear response formula for the Gor'kov Green function
and rewrite the result using the quasi-classical Green functions
and the evolution operator.\cite{aahn,higashitani_nagai} 
When we apply the formulation to the case with the constant order
parameter 
$\Delta_0, \Delta_1$, we obtain the susceptibility ( per unit volume )
$\chi_{zz}(0)$ at the surface
\begin{align}
 \frac{\chi_{zz}(0)-\chi_N}{\chi_N}& =\int_0^{\pi/2} d\theta\sin\theta
\pi T\sum_{n>0}f(\theta,\omega_n) \\
f(\theta,\omega_n)=&
\frac{\Delta_1^2\cos^2\theta}{(\omega_n^2+\Delta_0^2\sin^2\theta)
\sqrt{\omega_n^2+\Delta_0^2\sin^2\theta+\Delta_1^2
\cos^2\theta}},
\end{align} 
where $\chi_N$ is the normal state Pauli susceptibility and
$\omega_n$ is the Matsubara frequency.

In the presence of surfaces, the order parameter is  modified
cdas can be seen in Fig.~\ref{fig1}.\cite{roughp,vs} Near the specular
surface,
the perpendicular component $\Delta_1(z)$ is suppressed to zero,
while the parallel component $\Delta_0(z)$ is somewhat enhanced to compensate the
pairing energy.
In this report, we show the results of numerical calculations
using the self-consistent order parameter of Fig.~\ref{fig1}.
The details of the calculation
shall be reported elsewhere.

\begin{figure}[h]
\includegraphics[width=8cm]{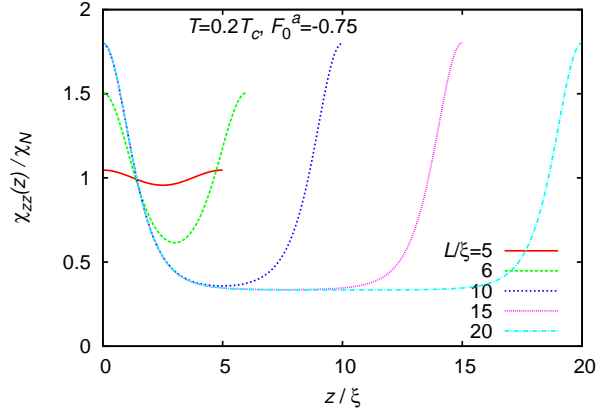}
\caption[ ]{Local distribution of $\chi_{zz}$ in ${}^3$He-B film
at a temperature $T=0.2T_c$. $\chi_{zz}$ is scaled by the
normal state value $\chi_N$.  Fermi liquid correction by $F_0^a$
 is taken into account. Film widths are the same as in Fig.~\ref{fig1}.
}
\label{fig2}
\end{figure}

\begin{figure}[h]
\includegraphics[width=8cm]{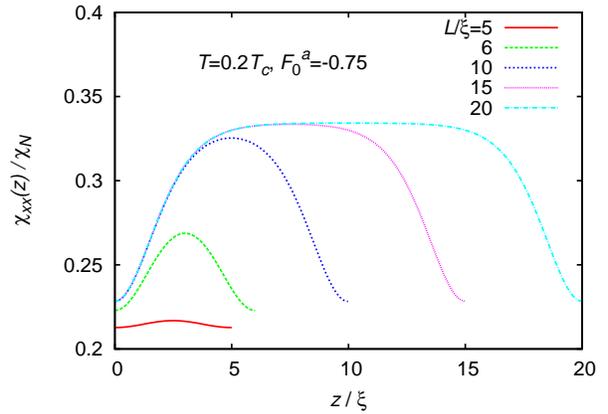}
\caption[ ]{Local distribution of $\chi_{xx}$ in ${}^3$He-B film
at a temperature $T=0.2T_c$. $\chi_{xx}$ is scaled by the
normal state value $\chi_N$. Fermi liquid correction by $F_0^a$
 is taken into account. Film widths are the same as in Fig.~\ref{fig1}.
Note that the vertical scale is different from Fig.~\ref{fig2}}
\label{fig3}
\end{figure}

In Fig.~\ref{fig2}, we show the local 
distribution of $\chi_{zz}(z)$ at a low temperature $T=0.2T_c$ 
for several width films. 
The Fermi liquid correction by $F_0^a$
is taken into account. The vertical axis is the local susceptibility
normalized by the normal state susceptibility. The enhancement of
the susceptibility can be clearly seen at the end surfaces. The bottom
value is almost equal to the bulk B-phase susceptibility.
In sufficiently thin films, the overall enhancement is found rather than
the surface enhancement. This is because the bound state
wave functions extend over the entire width of the film.
In constract to $\chi_{zz}$, the susceptibility $\chi_{xx}$ for the
magnetic field parallel to the surface is not enhanced at the surfaces
as can be seen in Fig.~\ref{fig3}. The surface value of $\chi_{xx}$ is
even smaller than the bulk susceptibility and is nearly equal to that
of the planar state with the Fermi liquid correction. 
These results clearly demonstrate that the surface bound states respond
to the magnetic field only in the direction of the surface normal.

\begin{figure}[h]
\includegraphics[width=8cm]{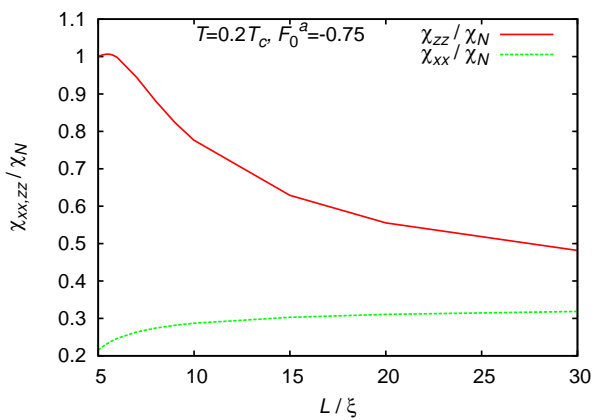}
\caption[ ]{Film width dependence of $\chi_{zz}$ and $\chi_{xx}$ 
 averaged over the film width at $T=0.2T_c$.}
\label{fig4}
\end{figure}

Finally we show in Fig.~\ref{fig4} the susceptibility $\chi_{zz}$
and $\chi_{xx}$
averaged over the film width. We can find that $\chi_{zz}$ even
exceeds the normal state Pauli susceptibility for sufficiently
thin films. In thicker films, $\chi_{zz}$ is still larger than
the B-phase bulk value, while $\chi_{xx}$ remains smaller. The
anisotropy is large enough to be observed. 

In conclusion, we have shown that the spin susceptibility of
superfluid ${}^3$He-B film has a strong anisotropy caused by
the surface bound states. The anisotropy is sufficiently large
to observe experimentally. We have considered, however, films
with specular end surfaces. In actual films on the substrate,
the surface scattering by the substrate will be diffusive.
One of the methods to avoid the diffuse scattering is to coat the
substrate by ${}^4$He layer. In fact, recent experiments of the
transverse acoustic impedance\cite{Saitoh,Murakawa} showed that
the specularity of the surface is considerably enhanced by the coating.
On the other hand, the susceptibility of the film with diffusive
surfaces is itself of interest. The density of
state at zero energy is known\cite{Zhang,roughp,vs,Volovik} to be increased
by the diffusive scattering, which might lead to further enhancement
of the susceptibility.

We thank Ryuji Nomura for bringing our attention to Refs. 19-24.
This work is supported in part by a Grant-in-Aid for Scientific Research 
Priority Area (No. 17071009) 
and a Grant-in-Aid for Scientific Research (No. 21540365) 
from the Ministry of Education, Culture, Sports, Science and 
Technology of Japan.

\end{document}